\input{epsf}

\documentclass[preprint,preprintnumbers,amsmath,amssymb]{revtex4}
\usepackage{graphicx}

\begin{document}
\draft
\title{Indirect constraints on the dark matter interpretation of excess 
positrons seen by AMS-02}
\author{Man Ho Chan}
\address{Department of Science and Environmental Studies, The Hong Kong 
Institute of Education \\ 
Tai Po, New Territories, Hong Kong, China}

\begin{abstract}
Recently, an excess of high energy positrons in our Galaxy has 
been observed by AMS-02. The spectrum obtained can be best fitted with the 
annihilation of $\sim$ TeV dark matter particles. However, recent 
analysis of Dwarf galaxies by Fermi/LAT observations highly 
constrains the TeV dark matter annihilation cross-section, and rules out 
the $b\bar{b}$ and all the leptophilic
channels except $4-\mu$ channel. In this article, I show that the 
remaining possible $4-\mu$ channel is also ruled out by using the 
observational data from cool-core clusters. Therefore, all the leptophilic 
channels that can account for the excess positrons seen in AMS-02, HEAT, 
and PAMELA are ruled out.
\end{abstract}

\maketitle
\subsection{Introduction}
Recently, the observations of high-energy positrons by HEAT 
\cite{Beatty}, PAMELA \cite{Adriani} and AMS-02 
\cite{Aguilar,Accardo,Aguilar2} reveal 
some excess emissions in our Galaxy. These 
excess emissions cannot be easily explained by standard astrophysical 
mechanisms \cite{Lopez}. Some studies propose that pulsars could 
generate enough high-energy positrons to account for 
the excess \cite{Hooper,Yuksel,Delahaye2,Profumo,Linden,Delahaye,Mauro}. 
On the other hand, many studies of the positron excess are now focusing on 
the annihilation of dark matter particles 
\cite{Delahaye3,Cirelli2,Bergstrom,Spanos,Ibarra,Cao}. Later, Boudaud et 
al. (2015) perform a detailed analysis on the latest AMS-02 
measurements and give a robust constraint on different annihilation 
channels such as $b\bar{b}$, $e^+e^-$, $W^+W^-$ and $\mu^+\mu^-$. They use 
more accurate data and look at a set of 1623 different 
combinations of the cosmic ray transportation parameters. They find that 
the allowed dark matter parameter space increases, and some other channels 
such as $4-\mu$ and $4-\tau$ can now provide excellent fits to the 
positron excess. The 
ranges of the best-fit annihilation cross-section $<\sigma v>$ and dark 
matter mass $m_{\chi}$ are $<\sigma v> \sim 
(10^{-24}-10^{-21})$ cm$^3$ s$^{-1}$ and $m_{\chi} \sim 0.1-10$ TeV 
respectively \cite{Boudaud}. The annihilation cross-sections obtained are 
larger than the 
expected one for a thermal relic. Such an enhancement could arise due to 
substructures in the dark matter distribution \cite{Lopez} or Sommerfeld 
enhancement \cite{Sommerfeld}.

Based on the result from \cite{Boudaud}, Lopez et al. (2015) use the 
Fermi/LAT gamma-ray data from dwarf galaxies to further constrain the 
cross-section and dark matter mass, especially for the 
$b\bar{b}$ and leptophilic annihilation 
channels. They find that all except the $4-\mu$ channel are ruled out 
\cite{Lopez}. Moreover, the multichannel combinations into $b\bar{b}$ 
and leptons are also excluded unless the branching ratios are allowed to 
deviate from their best-fit values \cite{Lopez}. The 
$4-\mu$ channel can escape from Fermi/LAT constraints because the emission 
of gamma-ray from dark matter annihilation is much less than the other 
channels. Most of the energy is given to the positron-electron pairs.

If the proposed TeV dark matter is really the dark matter in our universe, 
we may also constrain the dark matter properties by using the 
observational data in galaxy clusters. In this article, I show that the 
luminosity due to the cooling of the electron-positron pairs 
produced from dark matter annihilation through $4-\mu$ 
channel is larger than the observed luminosity in some nearby cool-cored 
galaxy clusters. In other words, all the leptophilic channels that can 
account for the positron excess are ruled out.
 
\subsection{Dark matter annihilation in galaxy clusters}
In the following, we consider dark matter annihilation through $4-\mu$ 
channel: $\chi \chi \rightarrow \phi \phi \rightarrow 4 \mu$, where 
$\phi$ is a mediator particle. It is so 
special because nearly 90\% of the energy from annihilation goes to the 
electron-positron pairs (only less than 1\% of the energy from 
annihilation contributes to gamma-ray). It is the reason why this 
channel can escape the bounds set by gamma-ray observations 
\cite{Lopez}.
The ranges of dark matter mass and cross-section that can account for the 
positron excess are $m_{\chi}=0.59 \pm 0.02$ TeV and $<\sigma v>=(5.87 \pm 
0.36) \times 10^{-24}$ cm$^3$ s$^{-1}$ respectively \cite{Boudaud} 
(based on the benchmark set of cosmic ray propagation model). The 
positron (or electron) 
spectrum $dN_e/dE$ for this channel has been computed in \cite{Cirelli} 
(see Fig.~1). The high-energy positrons produced would diffuse 
outward and 
cool down within the hot gas in galaxy clusters. The equilibrium positron 
spectrum can be calculated by using the diffusion equation \cite{Storm}:
\begin{equation}
\frac{\partial}{\partial t} \frac{dn_e}{dE}=\nabla \left[D(E,r) 
\nabla \frac{dn_e}{dE} \right]+ \frac{\partial}{\partial E} 
\left[b(E) \frac{dn_e}{dE} \right]+Q(E,r),
\end{equation}
where $dn_e/dE$ is the equilibrium electron/positron density spectrum, 
$D(E,r)$ is the spatial diffusion coefficient, 
$b(E)$ is the cooling rate and $Q(E,r)$ is the 
source term. Since the electron/positron radiation timescale is much 
shorter than the spatial diffusion timescale, we can neglect the time 
depedence term on the left hand side and the spatial dependence term on 
the right hand side of Eq.~(1). Therefore, the equilibrium density 
spectrum is \cite{Storm}:
\begin{equation}
\frac{dn_e}{dE}= \frac{<\sigma v> \rho_{\chi}^2}{2m_{\chi}^2b(E)} 
\int_E^{m_{\chi}}dE' \frac{dN_e}{dE'},
\end{equation}
where $\rho_{\chi}$ is the mass density of dark matter.

For electrons or positrons, there are four major ways of cooling: 
synchrotron radiation, inverse Compton scattering, Coulomb loss and 
bremsstrahlung \cite{Longair}. Since most of the electrons or positrons 
produced from the TeV dark matter annihilation through $4-\mu$ channel 
have energy above 1 GeV, the 
cooling rate would be dominated by synchrotron and inverse Compton 
sccattering. Therefore, we have \cite{Storm}
\begin{equation}
b(E) \approx \left[0.079 \left(\frac{E}{\rm 1~GeV} \right)^2 
\left( \frac{B}{\rm 1~\mu G} \right)^2+0.79 \left( 
\frac{E}{\rm 1~GeV} \right)^2 \right]~\rm eV/yr,
\end{equation}
where $B \sim 1$ $\rm \mu$G is the magnetic field in a typical galaxy 
cluster \cite{Storm}.

In equilibrium, the energy emitted by synchrotron radiation and inverse 
Compton scattering from electrons and positrons will finally leave the 
galaxy cluster and contribute to the total luminosity. The total 
luminosity due to the cooling of positrons and electrons within a radius 
$r$ in the cluster centre is 
\begin{equation}
L=2\int_0^r \int_0^{\infty}b(E) \frac{dn_e}{dE}dE(4 \pi r^2)dr.
\end{equation}
Assume that the density of dark matter is modelled by NFW profile 
$\rho_{\chi}=\rho_sr_s/r$ for $r \ll r_s$, where $\rho_s$ and 
$r_s$ are scale density and scale radius respectively \cite{Navarro}. The 
values of $\rho_s$ and $r_s$ can be calculated by the 
mass-concentration relation \cite{Schaller} and the virial radius 
obtained 
from x-ray observations \cite{Chen}. Therefore, by using Eq.~(2), Eq.~(4) 
can be written 
explicitly as 
\begin{equation}
L=4 \pi \frac{<\sigma v>}{m_{\chi}^2}\rho_s^2r_s^2r \int_0^{\infty} 
\int_E^{m_{\chi}}dE' \frac{dN_e}{dE'}dE.
\end{equation}
By using the energy spectrum in Fig.~1 and $m_{\chi}=0.59 \pm 0.02$ 
TeV, the integral in the above equation is equal to 556 GeV.

\begin{figure}
 \includegraphics[width=80mm]{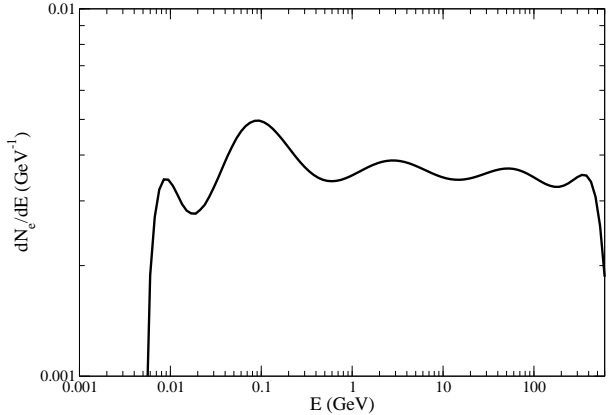}
 \caption{The energy spectrum of positron/electron produced per one 
annihilation through $4-\mu$ channel \cite{Cirelli}. Here, we use 
$m_{\chi}=0.59$ TeV.}
\end{figure}

\subsection{Compare with the observations}
In general, the luminosity due to positron and electron cooling is not 
very high ($L \sim 10^{39}$ erg s$^{-1}$ within 3 kpc). For a 
typical galaxy cluster with mean temperature $T=5$ keV, the luminosity 
of hot gas is about $L \sim 10^{40}$ erg s$^{-1}$ within 3 kpc. 
Nevertheless, observations reveal that some cool-cored clusters have low 
temperature ($T \sim 1$ keV) at the centre due to the strong cooling 
signatures \cite{Sanderson}. Therefore, the luminosity in 
the central part of cool-cored clusters is generally lower than that of a 
typical galaxy cluster. Provided that the luminosity due to positron and 
electron cooling should not exceed the observed luminosity, these 
cool-cored clusters can be good candidates to constrain the parameters of 
dark matter annihilation.

By using the observational data from \cite{Sanderson}, we can identify 
four good candidates (A262, A2199, A85 and NGC5044) to achieve our 
purpose. These clusters are chosen since they are characterized 
by a temperature of the hot gas that is less than 3 keV for $r<3$ kpc. 
Since the observed hot gas luminosity is assumed to be dominated by 
bremsstrahlung radiation (the radiation from recombination contributes 
less than 2\% of the luminosity), the observed luminosity is given by
\begin{equation}
L_o= \Lambda_0T^{1/2}\int_0^r 4 \pi r^2n^2dr,
\end{equation}
where $\Lambda_0=1.4 \times 10^{-27}$ erg cm$^3$ s$^{-1}$ and $n$ is the 
number density of hot gas. In general, the surface brightness profile 
of hot gas can be fitted with a single-$\beta$ model 
\citep{Cavaliere,Chen}
\begin{equation}
S(r)=S_0 \left(1+\frac{r^2}{r_c^2} \right)^{-3 \beta+1/2},
\end{equation}
where $S_0$ is central brightness, $r_c$ and $\beta$ are fitted 
parameters. This allows us to construct the radial gas number density 
distribution \citep{Reiprich}:
\begin{equation}
n=n_0 \left(1+ \frac{r^2}{r_c^2} \right)^{-3 \beta/2},
\end{equation}
where $n_0$ is the central number density. In fact, the surface
brightness profile of A85 is better described by a double-$\beta$
model \cite{Chen}. Nevertheless, the calculation of Eq.~(6) for A85 by 
using the double-$\beta$ model would just give a result with about 5\% 
smaller than 
that using the single-$\beta$ model. Therefore, for simplicity, we still 
use the single-$\beta$ model to calculate $L_o$. The observed 
temperature and the parameters used are summarized in Table 1.

By using $m_{\chi}=0.59 \pm 0.02$ TeV and $<\sigma v>=(5.87 \pm 0.36) 
\times 10^{-24}$ cm$^3$ s$^{-1}$, we can calculate the lower limit of $L$ 
by using Eq.~(5). On the other hand, by considering the uncertainties of 
the parameters (in Table 1), we can get the upper limit of $L_o$ by using 
Eq.~(6). In Table 2, we summarize the limits of $L$ and $L_o$ 
for A262, A2199, A85 and NGC5044. We can notice that $L>L_o$ in A262 and 
A2199 while $L<L_o$ in A85 and NGC5044. This suggests that the cooling 
rate due to 
positrons and electrons in A262 and A2199 is too high. The temperature 
should be much less than the observed one. To avoid 
exceeding the observed luminosity, the upper limit of the annihilation 
cross-section constrained by A262 is $<\sigma v><2 \times 10^{-24}$ cm$^3$ 
s$^{-1}$ for $m_{\chi}=0.59$ TeV, which is a few times smaller than the 
proposed one that can account for the excess positrons. In other words, 
the $4-\mu$ annihilation channel should be ruled out unless the 
dark matter cross-section and rest mass are significantly deviated 
from the best-fit values.

\begin{table}
\caption{The parameters used in our calculations \cite{Chen}. Here, $T$ 
is the temperature of the cool core in each cluster \cite{Sanderson}.}
 \label{table1}
 \begin{tabular}{@{}lcccc}
  \hline
   & A262 & A2199 & A85 & NGC5044\\
  \hline
  $\rho_s$($10^{14}M_{\odot}$ Mpc$^{-3}$) & 14.1 & 9.56 & 8.34 & 14.7 \\ 
  $r_s$(kpc) & 172 & 334 & 444 & 127 \\
  $\beta$ & $0.443^{+0.018}_{-0.017}$ & $0.665^{+0.019}_{-0.021}$ & 
$0.532^{+0.004}_{-0.004}$ & $0.524^{+0.002}_{-0.003}$ \\
  $r_c$(kpc) & $41^{+11}_{-9}$ & $139^{+10}_{-9}$ & $82^{+3}_{-3}$ & 
$11^{+0}_{-0}$ \\
  $n_0$($10^{-2}$ cm$^{-3}$) & $0.81^{+0.13}_{-0.09}$ & 
$0.83^{+0.03}_{-0.03}$ & $2.57^{+0.10}_{-0.10}$ & $3.45^{+0.03}_{-0.03}$ 
\\
  $T$(keV) & $<1$ (for $r<2$ kpc) & $<2$ (for $r<3$ kpc) & $<3$ (for $r<3$ 
kpc) & $<0.8$ (for $r<1$ kpc) \\
  \hline
 \end{tabular} 
\end{table}

\begin{table}
\caption{The luminosity due to positron/electron cooling $L$ and the 
observed luminosity of the cool core $L_o$. Here, we use $m_{\chi}=(0.59 
\pm 0.02)$ TeV and $<\sigma v>=(5.87 \pm 0.36) \times 10^{-24}$ cm$^3$ 
s$^{-1}$.}
 \label{table2}
 \begin{tabular}{@{}lcc}
  \hline
  Cluster & $L$(erg s$^{-1}$) & $L_o$(erg s$^{-1}$) \\
  \hline
  A262~~ & $~>8.3 \times 10^{38}~$ & $<4.2 \times 10^{38}$ \\
  A2199~~ & $~>2.2 \times 10^{39}~$ & $<1.7 \times 10^{39}$ \\
  A85~~ & $~>2.9 \times 10^{39}~$ & $<2.0 \times 10^{40}$ \\
  NGC5044 & $~>2.5 \times 10^{38}~$ & $<6.4 \times 10^{38}$ \\
  \hline
 \end{tabular} 
\end{table}

\subsection{Discussion}
In this article, we show that the cooling of the positrons and electrons 
produced from dark matter annihilation through $4-\mu$ channel can produce 
a significant amount of energy. This energy is a few times larger than the 
observed luminosity in two cool-cored galaxy clusters (A262 and A2199). It 
suggests that a dark matter annihilation through the $4-\mu$ channel with 
a cross-section sufficiently large to fit the high-energy positrons excess 
is not possible.

However, the $4-\mu$ channel is the only viable leptophilic channel to 
reconcile the tension between dark matter interpretation of excess 
positrons seen in AMS-02 and the gamma-ray constraint from dwarf galaxies. 
If our analysis is correct, all the leptophilic channels are ruled out by 
observations. Nevertheless, in the above calculations, we just have 2 
galaxy clusters to perform the analysis. More observational data from 
cool-cored clusters can provide a better verification in this issue.

Nevertheless, it is also possible to have other annihilation channels, 
such as quarks, vector and Higgs boson channels. The required dark matter 
mass is heavier than $\sim 10$ TeV, which would produce antiprotons at 
high energy. Recent analysis from \cite{Lin} suggests that dark matter 
annihilation through $b\bar{b}$ or $W^+W^-$ can provide good fits to the 
cosmic antiproton to proton ratio up to $\sim 450$ GeV. For example, 
for $W^+W^-$ channel, the required 
ranges of mass and cross-section of dark matter are $m_{\chi}=(4.4-36.3)$ 
TeV and $<\sigma v>=(3.7 \times 10^{-24}-3.5 \times 10^{-22})$ cm$^3$ 
s$^{-1}$ respectively \cite{Lin}. Obviously, the ranges are too wide 
to constrain the properties 
of dark matter. Moreover, the required cross-section is smaller than the 
one that can account for the excess positrons seen in AMS-02 ($<\sigma 
v>=(5.10 \pm 0.48) \times 10^{-22}$ cm$^3$ s$^{-1}$) \cite{Boudaud}. The 
same problem also applies to $b\bar{b}$ channel. Therefore, 
there is some inconsistency between the results from antiproton data and 
positron data. By combining all the results and constraints, it leaves 
only a small window for dark matter interpretation to resolve the 
conflicts among the positron data (AMS-02, HEAT, PAMELA), antiproton data 
(AMS-02), and gamma-ray data (Fermi/LAT).

\end{document}